# Framework for Risk-Based IoT Cybersecurity Audit Engagements

Danielle Hanson
Department of Computer Science
North Dakota State University
Fargo, ND, USA
danielle.jean.hanson@ndsu.edu

Jeremy Straub[1]
Center for Cybersecurity and AI
University of West Florida
Pensacola, FL, USA
jstraub@uwf.edu

**Abstract**
The use of Internet of Things (IoT) devices is growing at a rapid rate. While much of this growth is consumer devices, IoT devices are also commonly found in corporate and industrial environments, as well. These devices can be organization-owned and managed by an information technology unit, deployed organizationally without the knowledge and involvement of technology staff or brought in to the corporate environment by user-owners. In each case, these devices may have access to corporate networks and data and are, thus, important to consider as part of organizational cybersecurity risk assessment. Despite the prevalence of these devices, there is little literature about how to audit their security. This paper presents a risk-based auditing framework which can be used by both internal and external auditors, of any experience level and in any industry, to assess IoT devices.

**Keywords:** internet of things, IoT, cybersecurity, risk-based audit, small devices, internet-connected, framework, assessment

## 1. Introduction

Internet of Things (IoT) devices, also known as connected devices or smart devices, are physical devices that have the ability to collect and share data with other devices over communication networks. IoT excludes traditional computing devices such as mainframe, desktop and laptop computers, smartphones, and servers.

There have been numerous attacks targeting IoT devices, such as the 2024 Androxgh0st botnet attacks that targeted IoT devices and stole cloud credentials [1]. Given this, the over 16 billion IoT devices worldwide [2]—many of them used in critical infrastructure sectors such as healthcare, energy, and manufacturing—should be as much of (or perhaps even more of) a concern to organizations as traditional computing devices are.

The lack of historical auditing of IoT devices, the lack of standards and the lack of regulation may result in organizations failing to fully consider the security of IoT devices. Some organizations may consider them as part of their IT security, missing key differentiating factors of IoT devices. Other organizations may fail to consider them at all or fail to consider a subset of devices, such as those connected to organization networks by user-owners.

IoT security legislation is new and differs from jurisdiction to jurisdiction. One example of IoT security legislation, that applies to manufacturers or sellers of IoT devices in the United States, is the IoT Cybersecurity Improvement Act of 2020 [3]. This law applies to devices procured or used by federal agencies. It requires them to follow standards created by the National Institute of Standards and Technology (NIST).

Another example is state laws such as California's SB-327 Information Privacy: Connected Devices [4] and Oregon's HB-2395, which amended ORS 646.607 [5]. Both laws require IoT device manufacturers to add basic security features to their devices but have limited mechanisms for enforcement.

---

[1] This work was partially completed while J. Straub was at North Dakota State University.

There is no law specifically regulating IoT device security for general consumers at the federal level in the United States. This leaves the security of many consumers – and most organization's IoT networks – entirely in their own hands. Because of this, organizations that have a need to audit their IoT device and network security, but are not specifically required to follow existing security and privacy regulation—such as the National Institute of Standards and Technology Special Publication 800-171 (NIST SP 800-171) [6], can benefit from a risk-based audit approach. A risk-based approach tailors an audit to specific organizational needs and directs resources to where they are most needed. This approach maximizes benefit, in the absence of a regulatory framework specifying specific checks and requirements.

This paper proposes an IoT cybersecurity audit framework. It continues, in Section II, by providing background on IoT auditing, existing frameworks, and types of audits. Then, it presents a risk-based framework for auditing IoT security. The goal of this framework is to aid auditors of IoT devices and networks and to help auditors perform an effective and efficient risk-based audit, with consistent, understandable, and useful reporting. After presenting the framework, the paper provides examples of applications the framework may have outside of IoT security auditing. Finally, it concludes with a discussion of future work that could further enhance the framework.

## 2. Background

This section reviews prior work in several areas which can inform the development of an IoT cybersecurity audit framework. It is crucial to note that there are a number of key differences between traditional security audits and IoT security audits. Traditional computing devices frequently come with built-in security features and options and are the focus of many cybersecurity efforts and trainings. IoT devices are numerous, but their security may not be in the front of the average worker's mind. They are also frequently manufactured or configured with no or limited security features. Their vulnerabilities cannot always be patched.

Additionally, because they are resource-constrained compared to traditional computing devices, they present particular challenges with regards to encryption and other security technologies. Many regulations that influence security protocols, such as the Health Insurance Portability and Accountability Act (HIPAA) [7], apply to IoT devices; however, because they were developed before IoT devices became widespread, they do not address them specifically. Thus, medical IoT security, privacy, and HIPAA compliance is now a concern [8].

These characteristics create extra challenges with the identification of IoT audit areas, designing IoT audit procedures, and recommending remediations for IoT security vulnerabilities, as compared to traditional security audits. Additionally, because of the prevalence of IoT devices in many organizations, it may be necessary to incorporate IoT auditing into many audits that previously included only – or primarily – traditional computing devices.

Whether an IoT audit and a traditional security audit are performed by the same team or by different, specialized teams depends on an auditee's size and the technical expertise of the audit team. An IoT security auditing framework can help auditors who are inexperienced with IoT auditing perform high-quality engagements without requiring specialist help.

### 2.1. IoT Security Assurance

There is some existing research on the topic of IoT security assurance. However, as shown in [9], there are few security auditing frameworks specific to IoT. One of these existing frameworks was proposed in [10]. It described a risk assessment methodology and model which considered the IoT system elements of assets, vulnerabilities, and threats. Bena, Bondaruc and Polimeno [11] proposed a framework for an edge device microservices architecture for providing security assurance to IoT systems. In [12], dynamic risk assessments for medical IoT – which are risk assessments that update continuously rather than periodically and can help auditors continue to use relevant testing procedures during longer engagements – were discussed. The potential usefulness of dynamic risk assessment for IoT security assurance is not limited to medical IoT.

Boye, Kearney and Josephs [13] focused on risk assessment, with an emphasis on continuous risk assessment for critical infrastructure industrial IoT, which may require special considerations when undergoing security assessment. It may not be able to be subjected to – or even risk – the same interruptions that other IoT networks can, lest important services be disrupted. In [14], a risk-based framework focused on the cybersecurity certification of individual IoT devices. rather than IoT networks as a whole, was proposed. This is potentially useful for enabling procurers of IoT devices to make informed decisions. Additionally, components of this research on risk-based automated assessment of individual IoT device security can potentially overlap with the testing of IoT network security.

The collection of existing frameworks demonstrates that there is a need for IoT security auditing frameworks and a particular need to those that incorporate risk assessment. Several of these existing frameworks are industry-specific, but IoT is used across industries.

*2.2. Auditing*

Audits can be categorized as being risk-based or compliance-based. Risk-based audits are designed based on a risk assessment of the auditee and tailored to the auditee's circumstances. Compliance-based audits are designed to verify compliance with a specific standard, such as HIPAA or a Cybersecurity Maturity Model Certification (CMMC) [15] level.

Compliance-based auditing is used when there is a standard that provides a checklist that must be followed by the auditee. It may be required in these circumstances. Risk-based auditing can be used for any auditee and is well-suited for ad hoc audits, when standard requirements are not robust and detailed, or when relevant standards do not exist.

Although this framework is designed specifically for risk-based audits, it can also be used for audits with compliance-based elements. For example, it might be used for a CMMC audit, which requires verification of compliance with various NIST standards. This is because, even in the presence of a checklist that must be followed, efficiently addressing all standard requirements still involves risk assessment, prioritization of resources, and flexibility. Organizations which have the responsibility to comply with existing security standards that apply to their IoT networks can still benefit from risk-based auditing as a tool for identifying security gaps that a given security standard does not address.

## 3. Audit Activities Overview

IoT security audit activities can be categorized into three phases: planning, fieldwork, and post-fieldwork. This is depicted in Figure 1.

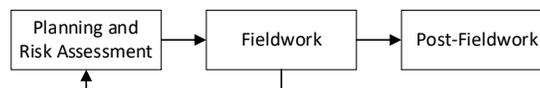

Fig. 1. Iterative risk assessment process in a risk-based audit.

During the planning phase, the engagement scope is defined and the auditor gathers preliminary information, conducts an initial risk assessment, and creates an audit plan. During the fieldwork phase, the auditor conducts information gathering and testing procedures. Finally, during the post-fieldwork phase, the auditor's work is reviewed, the audit report is drafted, presented to auditee management and governance (or other applicable persons), and finalized. The goal of an IoT audit is to provide an opinion on the security of the auditee's IoT networks and devices in a report that is useful and understandable to the auditee's stakeholders.

IoT auditors should follow standard professional practices. They should avoid situations in which there is or appears to be a conflict of interest. These situations include—but are not limited to—when the auditor sells IoT devices, has designed or configured all or a portion of the auditee's IoT devices and network, has an ownership stake in the auditee, is related to or has a close business relationship with key auditee personnel, and/or if the auditor is considering a position at the auditee organization.

IoT auditors should not accept non-trivial or monetary gifts from the auditee or its personnel or allow domineering, pushy, or threatening auditee personnel to affect their judgment. An auditor may need to bring up behaviors with superiors, address them with auditee governance, or withdraw from an engagement, depending on issue severity.

The following three sections discuss the three audit phases in more detail.

## 4. Planning

The IoT audit planning process includes the following activities:

1. Gain an understanding of the environment, both external (economic, physical, political, regulatory, etc.) and internal, in which the auditee conducts operations.
2. Evaluate what security requirements, if any, the auditee is subject to. Security requirements can come from sources including laws, professional associations, industry standards, certifications, and insurers.
3. Gather information about the auditee's IoT infrastructure and documented cybersecurity controls and policy. This includes performing interviews and testing, as needed, to verify the accuracy of this information.
4. Conduct an initial risk assessment of the auditee, involving the entire engagement team, based on the information gathered in steps 1-3.
5. Determine the nature, timing, and extent of audit procedures that are needed, based on the auditee's IoT cybersecurity risks and create an appropriate budget with the understanding that procedures may change as the audit proceeds.
6. Revise the risk assessment and audit plan throughout the audit, as needed, making sure to document all changes.

Risk assessment is an iterative process. An audit plan can and should change throughout the engagement in response to previously unknown discoveries. This may be particularly true with IoT audits, where auditors may discover additional types, sources and uses of IoT devices during an engagement.

It is important to conduct an initial risk assessment and create an audit plan before testing begins. Everyone who is significantly involved in the engagement should participate in the initial risk assessment meeting, as it is helpful to have multiple perspectives when considering the potential issues. Meeting participation can provide a training opportunity for inexperienced auditors.

Internal and external risks and the conditions the auditee is or may be subject to should be considered. Examples of these include the auditee's regulatory environment, the expertise of their information technology staff and the staff that install and configure IoT devices and networks, the presence and expertise of security personnel, management's attitudes toward internal controls, industry standards, how IoT is used within the organization, and applicable IoT laws.

When considering possible cybersecurity threats to IoT, auditors can use resources such as the MITRE ATT&CK matrix for enterprises [16], which categorizes techniques used by cyber-attackers into fourteen tactics: "reconnaissance", "resource development", "initial access", "execution", "persistence", "privilege escalation", "defense evasion", "credential access", "discovery", "lateral movemen", "collection", "command and control", "exfiltration", and "impact". Each of these tactics, or the equivalent, if using another attack framework (such as the MITRE ATT&CK matrix for ICS [17]) should be considered during the risk assessment and when designing testing procedures.

Auditors can also use internal control frameworks, such as the Committee of Sponsoring Organizations of the Treadway Commission (COSO) Internal Control-Integrated Framework [18] and the NIST Cybersecurity Framework (CSF) 2.0 [19], as resources. These frameworks may be helpful for anticipating where internal control weaknesses are likely to occur.

The ultimate goal of risk assessment is to reduce audit risk – the risk of the auditor expressing an inappropriately favorable opinion on the auditee's IoT security – while maintaining audit efficiency. This is based on the Public Company Accounting Oversight Board's definition of audit risk in a financial statement audit as "the risk that the auditor expresses an inappropriate audit opinion when the financial statements are materially misstated" [20].

Thorough planning and a well-thought-out and documented risk assessment, that is updated throughout the audit engagement, are necessary for a successful risk-based audit. Risk assessments should be updated for every engagement, even when the auditor has worked with the auditee before, as changes in technology, policy, personnel, political environment, and regulation can affect the auditee's risk level.

**5. Fieldwork**

Audit procedures conducted during fieldwork can be broadly categorized as either information gathering or testing procedures. Information gathering procedures inform testing procedures, which determine the results of the audit. At times, testing procedures can reveal the need for more information gathering, such as if it was discovered that IoT network documentation is out of date, there are weaknesses in internal controls, or employees are not following documented policies. Both types of procedures are now discussed.

*5.1. Information Gathering Procedures*

Information gathering procedures – which include interviewing, observation, and inspection – are an important part of the audit planning and audit testing processes. The quality of information obtained is as important as the quantity and types of information obtained. It is, thus, important to watch for inconsistent, unclear, or outdated information, which can be misleading and waste time or reduce audit quality. For example, when receiving network or facilities maps, the auditor should ask how often they are updated and what the update process involves. The auditor should also verify a portion of the maps to test for undocumented changes. When receiving policy documents, the auditor should ask how policies are made and how policies and changes are communicated to employees. They should also interview individuals responsible for following the policies to see if their processes may differ from the documentation.

The results of the information gathering procedures should be documented. This includes documenting oral conversations and interviews. When oral conversations and interviews are used for gathering information to use in an audit, the auditor should document who the conversation or interview was with, the date of the conversation or interview, and the key content of the conversation or interview.

Interviewees should not feel interrogated or like they need to be guarded. Auditors should explain the purpose of the audit and interview and adjust their language and use of technical jargon to the appropriate level for the interviewee.

Employees might also not know what IoT devices are or realize the extent to which they are used within the auditee organization. It may be necessary to define IoT for non-IT personnel who may use IoT devices without knowing that they are doing so. During interviews, auditors should remain alert for any discrepancies between procedure documentation and how work is actually being performed. Auditors can also observe work being performed and inspect facilities and IoT devices when gathering information about the auditee. They should note where observation and inspection differ from information gathered from interviews or policy documents.

*5.2. Testing Procedures*

Testing procedures are designed by the auditor to fit the security risk of the auditee and the scope of the audit. Each procedure should be designed in response to a specific risk. Each procedure should be documented, along with the risk it addresses, the results of the procedure, and any necessary follow-up activities to the procedure. While surprises can happen in any audit, planning and risk assessment – before testing procedures begin – help ensure the audit is effective, while progressing in a timely manner.

As the audit progresses, it may be necessary to adjust, expand, add, or eliminate testing procedures. When an auditor discovers unknown information about the auditee's IoT infrastructure, they should consider the following questions:

1. How does this affect the risk to the confidentiality and integrity of the auditee's data—both in transit and at rest?
2. How does this information affect the operational security of the auditee and the auditee's continuity of operations?

3. Does this have a regulatory impact on the auditee? IoT devices can collect, transmit, or store data that is subject to regulation.
4. How does this change the risk assessment of the auditee when combined with other known information?
5. How should the audit team respond to this information?

After going through the above questions and analyzing the results, the auditor should document the course of action they take and their reasoning.

*5.3. Automated Procedures*

The quantity of IoT devices used by many organizations makes IoT security audits a good candidate for using automated testing tools. One example tool is the software for operations and network attack results review (SONARR) [22] which, when given a representative model of a network, can examine the network and identify possible vulnerabilities. These possible vulnerabilities can then be investigated by the auditor. By using a representative model of the network, rather than the network itself, SONARR reduces the likelihood of business disruption from security testing, which can be a significant concern for complex mission critical systems that cannot be put at risk [22].

Using automated tools like SONARR can free auditor time and resources for other tasks and can allow auditors to perform testing procedures that would otherwise be too tedious or impractical to perform, within the timeframe of an engagement. SONARR incorporates a verifier feature that helps ensure the representative model accurately reflects the auditee's actual network [21]. This feature makes this tool especially useful for internal auditors and others who work repeatedly with the same systems and conduct follow-up testing after the initial audit is concluded.

Notably, a wide variety of tools can be leveraged by IoT auditors which may provide similar benefits and functionality as SONARR.

**6. Post-Fieldwork**

All auditor work and results should be documented with enough thoroughness that a knowledgeable professional, uninvolved with the engagement, can understand the nature, timing, and extent of testing performed and the reasoning behind the auditor's determinations [23]. As is standard best practice, disagreements between auditors and between auditors and management should be documented, and all documentation should be dated. All work and documentation should be reviewed for quality control, understandability, and thoroughness. Thorough audit documentation provides a resource for future engagements, encourages quality work, and provides justification for the auditor's conclusions.

Findings should be evaluated using the iterative decision making model described below:

- Step 1: Identify a control weakness.
- Step 2: Assess the likelihood and means of exploitation of the control weakness.
- Step 3: Evaluate the potential losses (financial and non-financial) faced by the organization and its stakeholders if the control weakness is exploited, as well as the regulatory impact of the control weakness.
- Step 4: Consider the existence of alternate or compensating controls that reduce or eliminate the likely and potential harm caused by the exploitation of the control weakness. Also, consider the existence of other weaknesses in security controls that are compounded by this control weakness.
- Step 5: Determine the overall severity of the control weakness and the warranted response. Consider whether or not it is possible for the client to remediate the finding while maintaining its existing IoT equipment.

When determining the severity of a control weakness and the adequacy of any compensating controls, auditors can use frameworks such as Lockheed Martin's Cyber Kill Chain [24], which lays out the seven stages to cyber-attacks carried out by external threat actors: "reconnaissance", "weaponization", "delivery", "exploitation", "installation", "command and control", and "actions on objectives". Controls at any stage of the Cyber Kill Chain can stop a specific attack or force an attacker to look for an alternative strategy. This makes the Cyber Kill Chain and similar models useful tools for evaluating the ability of compensating controls to reduce the impact of a weakness in IoT security controls.

Irrespective of the severity of the finding, auditors should provide a recommended response to the auditee. The auditor can also recommend alternative responses when multiple remediation methods are available. When practical, an estimated cost of remediating the finding may also be included. However, to preserve the auditors' independence, the auditor should not implement a recommendation for the auditee or sell them a security solution.

The auditor's final opinion on the auditee's IoT security, findings, and recommendations should be documented in a final report, which is presented to the auditee's governance (or the group that requested the audit). The auditor can provide management a list of audit findings, before presenting the report to governance, to give them an opportunity to correct any misunderstandings.

The responsibility for determining what is and is not a finding worth including in the report lies with the auditor, not with management. The auditor should not subordinate their judgment to managements' [25].

The final report from an IoT audit will typically begin with a signed and dated opinion letter, which summarizes the auditor's opinion of the auditee's internal controls over IoT security. If the auditor was unable to gather enough information to form an opinion of the auditee's internal controls over IoT security, they can note that they disclaim an opinion.

The report should include information about the auditee, their industry, the relevant regulations and standards they are subject to, and the ways they use IoT devices. The report should then list the audit findings and recommended remediations. The report should also indicate the severity of and risk associated with each audit finding.

**7. Other Uses of This Framework**

This framework is designed for use by internal and external auditors of IoT security. It can also be used by IT and security teams, with some modifications. IT and security teams can ignore the inapplicable conflict of interest and independence components. They should also alter the reporting to fit their internal needs and use the results of the assessment for planning, prioritizing, and budgeting their work.

The planning portion of the framework can be used for risk assessment by internal staff. The framework can also be used by educators for teaching students about IoT security auditing and the auditing and risk assessment processes, as it is designed to be understandable, concise, and useful as a training tool.

**8. Future Work and Conclusion**

This paper has proposed a preliminary IoT auditing framework. Future potential work in this area includes interviewing and surveying security auditors, of various skill levels and from various industries, to gain a better understanding of the challenges they face and how a framework can help address them.

Other potential future work includes the assessment of the framework by conducting cybersecurity audits and assessments using it, to evaluate its efficacy and practical usefulness. This will involve interviewing IT and security personnel who work for auditees to identify how their experience during IoT security audits can be improved. Future work can also focus on identifying how the framework can incorporate emerging technologies that can increase IoT audit efficiency and identifying how to maximize the benefit of IoT audits to stakeholders.

IoT audit is an increasingly important field where a great deal of research is needed. IoT auditors must navigate changing technologies and regulations while providing quality work and useful reporting. Planning and risk assessment, the appropriate use of automated tools, and useful and understandable reporting are all key to the risk-based IoT security auditing process. This risk-based IoT security auditing framework can serve as an aid for IoT auditors, at any level of seniority, when planning and conducting a risk-based audit of IoT security controls.

**References**

[1] F. Okeke, "How Androxgh0st Became 2024's Worst Malware," Technopedia, 1 January 2025. [Online]. Available: https://www.technopedia.com/andoxgh0st-malware. [Accessed 3 January 2025].
[2] S. Sinha, "State of IoT 2024: Number of connected IoT devices growing 13% to 18.8 globally," IoT Analytics, 3 September 2024. [Online]. Available: https://iot-analytics.com/number-connected-iot-devices/. [Accessed 3 January 2025].
[3] 116th Congress, "H.R.1668 - IoT Cybersecurity Improvement Act of 2020," 12 April 2020. [Online]. Available: https://www.congress.gov/bill/116th-congress/house-bill/1668.
[4] "SB-327 Information Privacy: Connected Devices," 28 September 2018. [Online]. Available: https://leginfo.legislature.ca.gov/faces/billTextClient.xhtml?bill_id=201720180SB327.
[5] 80th Oregon Legislative Assembly, "Enrolled House Bill 2395," 2019.
[6] NIST, "Protecting Controlled Unclassified Information in Nonfederal Systems and Organizations," 2024.
[7] 104th Congress, "H.R.3103 - Health Insurance Portability and Accountability Act of 1996," 1996.
[8] R. Kumar and R. Tripathi, "Towards design and implementation of security and privacy framework for Internet of Medical Things (IoMT) by leveraging blockchain and IPFS technology," The Journal of Supercomputing, pp. 7916-7955, 2021.
[9] D. Hanson and J. Straub, "A Systematic Review of Cybersecurity Audit Frameworks for the Internet of Things," 2024 IEEE International Conference on Cybersecurity and Resilience, 2024.
[10] W. Kang, J. Deng, P. Zhu, X. Liu, W. Zhao and Z. Hang, "Multi-dimensional Security Risk Assessment Model Based on Three Elements in the IoT System," 2020 IEEE/CIC International Conference on Communications in China, 2020.
[11] N. Bena, R. Bondaruc and A. Polimeno, "Security Assurance in Modern IoT Systems," 2022 IEEE 95th Vehicular Technology Conference, 2022.
[12] R. M. Czekster, P. Grace, C. Marcon, F. Hessel and S. C. Cazella, "Challenges and Opportunities for Conducting Dynamic Risk Assessments in Medical IoT," Applied Sciences, 2023.
[13] C. A. Boye, P. Kearney and M. Josephs, "Cyber-Risks in the Industrial Internet of Things (IIoT): Towards a Method for Continuous Assessment," International Conference on Information Security, pp. 502-519, 2018.
[14] S. N. Matheu-Garcia, J. L. Hernandez-Ramos, A. F. Skarmeta and G. Baldini, "Risk-based automated assessment and testing for the cybersecurity certification and labelling of IoT devices," Computer Standards and Interfaces, pp. 64-83, 2019.
[15] Defense Department, "Cybersecurity Maturity Model Certification (CMMC) Program," 2024.
[16] MITRE, "MITRE ATT&CK Enterprise Matrix," [Online]. Available: https://attack.mitre.org/matrices/enterprise/. [Accessed 8 January 2025].
[17] MITRE, "MITRE ATT&CK ICS Matrix," [Online]. Available: https://attack.mitre.org/matrices/ics/. [Accessed 8 January 2025].
[18] COSO, "Internal Control - Integrated Framework," [Online]. Available: https://www.coso.org/guidance-on-ic. [Accessed 8 January 2025].
[19] NIST, "The NIST Cybersecurity Framework (CSF) 2.0," 2024.
[20] PCAOB, "AS 1101: Audit Risk," PCAOB, [Online]. Available: https://pcaobus.org/oversight/standards/auditing-standards/details/AS1101. [Accessed 3 January 2025].
[21] J. Milbrath and J. Straub, "Incorporation of Verifier Functionality in the Software for Operations and Network Attack Results Review and the Autonomous Penetration Testing System," 13 September 2024. [Online]. Available: https://arxiv.org/abs/2409.09174. [Accessed 3 January 2025].
[22] M. Tassava, C. Kolodjski and J. Straub, "Development of a System Vulnerability Analysis Tool for Assessment of Complex Mission Critical Systems," arXiv, 2023.
[23] AICPA, "AU Section 339.10 - Audit Documentation".
[24] Lockheed Martin, "The Cyber Kill Chain," [Online]. Available: https://www.lockheedmartin.com/en-us/capabilities/cyber/cyber-kill-chain.html. [Accessed 8 January 2025].
[25] AICPA, "Code of Professional Conduct and Bylaws - 2.100.001".